\documentclass[11pt]{article}
\usepackage[english]{babel}
\usepackage{graphicx}
\input{epsf}

\usepackage[dvipsnames]{color}

\def\be{\begin{equation}}
\def\ee{\end{equation}}
\def\ba{\begin{eqnarray}}
\def\ea{\end{eqnarray}}

\def\12{{1\over 2}}

\def\ltsima{$\; \buildrel < \over \sim \;$}
\def\simlt{\lower.5ex\hbox{\ltsima}}
\def\gtsima{$\; \buildrel > \over \sim \;$}
\def\simgt{\lower.5ex\hbox{\gtsima}}

\begin{document}
\selectlanguage{english}

\title{\bf The Multiplicity of Main Sequence Turnoffs in Globular Clusters}
\author{M.~V.~ Ryabova$^1$ and Yu.~A.~Shchekinov$^{1,2}$\thanks{yus@phys.rsu.ru}\\
{\it $^1$Southern Federal University, Rostov-on-Don, Russia} \\
{\it $^{2}$Special Astrophysical Observatory,
 Nizhnyi Arkhyz, Karachai-Cherkessia, Russia} \\
{\small Received; in final form, June 22, 2007}}

\date{\small published in 2008, Astr. Repts, 52, 352;
originally in russian in Astr. Zh.,
85, p. 398}

\maketitle

\begin{abstract}
We present color-magnitude diagrams of globular clusters for models
with self-enrichment and pre-enrichment. The models with
self-enrichment turn out to have two or more main sequence turnoff
points in the color-magnitude diagram if the fraction of mass lost
by the globular cluster under supernova explosions does not exceed
95--97\%. The models with pre-enrichment can have only one main
sequence turnoff point. We argue that the cluster $\omega$ Cen
evolved according to a self-enrichment scenario.
\end{abstract}


\section{INTRODUCTION}

\noindent

Globular clusters (GCs) belong to the oldest stellar population of
the Galaxy. The ages of the oldest GCs (13–-14 Gyr) suggest that
they have formed as distinct structures at the very beginning of the
Galaxy formation, possibly preceding formation of the field
stellar population. Therefore, it is natural to assume that
heavy elements in GCs have been synthesized during the process of
stellar evolution
in the GCs themselves, and were not brought in with the gaseous
clouds from which the clusters formed. This point of view is
discussed recently as the self-enrichment hypothesis, contrary to the
pre-enrichment hypothesis, which assumes that the abundances of
metals currently
observed in GCs are essentially the same as their values in the gaseous
clouds of proto-GCs; in other words that the metals have been produced by  an external
stellar population~\cite{pre}. Note that the pre-enrichment
scenario seems to be required for at least some GCs (such as the young  group of GCs). This
follows from the bimodality of GCs, i.e. from the fact that they form
two groups with different metallicities, as was first noted
in~\cite{suchkov} (for later discussions see~\cite{gc2}). We stress
that the concept of self-enrichment does not exclude a possibility
that proto-GCs gaseous clouds have already contained some metals
produced by the very first Population III stars, though it
suggests that their amount was negligible
compared to the observed values.

Neither of these hypotheses can currently be rejected or accepted
confidently based on observations. Arguments exploiting expected
mass-metallicity correlation~\cite{MZcor} cannot be applied to GCs
directly due to their low binding energy~\cite{parm1}. Furthermore,
some evolutionary scenarios for GCs predict an anticorrelaton
between their mass and metallicity~\cite{MZcor,AJ}: $Z\propto
M^{-1}$. A weak anticorrelation between mass and metallicity
discovered in~\cite{parm1} for old GCs might argue in favor of the
self-enrichment hypothesis, but is not entirely convincing due to
the small sample studied ($\sqrt{N}\simeq 4.5$); in addition, this
anticorrelation is unstable to effects related to the decrease in
the masses of GCs due to evaporation~\cite{henon}, gravitational
encounters~\cite{spitzer}, and tidal forces~\cite{surdin}. Under
these circumstances, it seems necessary to search for possible
manifestations of self-enrichment in the color characteristics of
some particular GCs, and for possible expected differences between the
characteristics of stars in old and young GCs.

In this paper, we present the results of modeling of the fine structure
of the color-magnitude diagrams of GCs in the vicinity of their
main sequence turnoffs, and describe the dependence of this structure
on enrichment scenarios. Recently published detailed observational
studies of the GCs $\omega$ Cen~\cite{Helium} and NGC
2808~\cite{n2808} provide information about the star formation
history (SFH) and enrichment of these GCs. A comparison of our
modeling results with the observed color-magnitude diagrams for
$\omega$ Cen and NGC 2808 provides support for the self-enrichment
scenario for both $\omega$ Cen and NGC 2808.

$\omega$ Cen is distinguished among GCs in its unusually
high mass ($M\approx 3\cdot 10^6 M_\odot$) and the large scatter of
stellar metallicities~\cite{tsuj}. The latter seems to
provide evidence in favor of self-enrichment in this
cluster~\cite{sunt}. Furthermore, as the binding energy of
$\omega$ Cen does not substantially exceed the binding energies of
most GCs in the Galaxy~\cite{gne}, this suggests that at least some
fraction of the Galactic GCs were enriched in metals by nucleosynthesis
occurring inside the clusters themselves. The absence of stars with metallicities $[Z]<-2$ in
$\omega$ Cen does not necessarily require pre-enrichment to
this metallicity as noted in~\cite{tsuj}, but may provide evidence for triggering of star
formation in $\omega$ Cen by activity of supernovae in the cluster
(see~\cite{AJ} for discussion).

\section{DESCRIPTION OF THE MODEL}

\noindent

The parameters of the GCs were computed using a single-zone
model~\cite{AJ} based on the standard system of equations describing
evolution of the gas mass and the elemental abundances, and a
single-zone imitation of dynamics
associated with the energy released by
supernovae~\cite{matt,firman,shust,ciapp}.

\subsection{Chemistry}
The first model equation is the conservation of the gas mass:

\begin{equation}
  \label{eq1}
  \frac{dM_g}{dt} = -\psi\left(t\right) +
    \int\limits_{M_{min}}^{M_{max}}
      {\psi\left(t-\tau_{_M}\right) \varphi\left(M,t-\tau_{_M}\right) \left(M-M_r\right) dM} -
    \dot{M}_g^{out} + \dot{M}_g^{in},
\end{equation}
where $M_g$ is the mass of gas in the system, $M_r(M)$,
the mass of stellar remnants, $\psi(t)$, the
star formation rate (SFR), $\varphi(M,t)$, the initial
mass function (IMF) at time $t$, $M_{min}$ and $M_{max}$ are the minimum
and maximum masses of new-formed stars, $\tau_{_M}$ is the lifetime of
a star with mass $M$, and $\dot{M}_g^{out}$ and $\dot{M}_g^{in}$ are the
rates of ejection and accretion of gas by the system, respectively.

We used the standard Schmidt law~\cite{Vibe,shust} for the SFR:

\begin{equation}
  \psi\left(t\right) = f \rho^n V,
\end{equation}
where $f$ is the star formation efficiency (SFE), $\rho$, the average
density of the gas, and $V$, the volume of the stellar system.
Schmidt~\cite{schmidt} adopted the value $n = 2$. The quadratic
dependence follows from star formation models with self-regulation,
in particular, the model in which star formation is regulated by the
ionization of gas by UV photons from massive stars~\cite{cox}. The
efficiency $f$ is determined by the specific mechanisms providing
a power-law dependence of the SFR, $\psi\sim\rho^n$, and in
general determination of $f$ is a separate problem. A good estimate for
the SFE in the Schmidt law is probably provided by $f=2\cdot
10^7$~sm$^3$g$^{-1}$s$^{-1}$, as adopted in~\cite{Vibe,shust}.

We used a Salpeter IMF~\cite{Salpiter}:
\begin{equation}
    \varphi\left(M\right) \sim M^{-2.35}
\end{equation}
with the masses in the range from
$M_{min}$ to $M_{max}$.

The equation describing time variations of the
radius of the gaseous component $R_g$ is

\begin{equation}
\label{eq2}
  \frac{dR_g}{dt}=\frac{R_\ast^3 \gamma\varepsilon_0 R_{_{SN}}\left(t\right)}
  {3 G M_{_{GC}} M_g R_g}-\frac{R_g}{6 \tau_d},
\end{equation}
where $M_{_{GC}}$ is the total mass of the GC, $R_\ast$ is the radius of
the stellar component, while $R_g$ is the gas component radius,
$G$, the gravitational constant, $\gamma$, the efficiency of the
transformation of the supernova explosion energy into the energy
of turbulent gas motions, $\varepsilon_0$, the explosion energy,
$\tau_d$, the timescale for energy dissipation
in cloud collisions, and $R_{_{SN}}(t)$, the rate of type
II supernovae. Equation (\ref{eq2}) is a spherically
symmetric version of the dynamical equation written fist
by~\cite{firman} for a flat geometry.

The type II supernovae rate $R_{_{SN}}(t)$ is defined by
the integral
\begin{equation}
  R_{_{SN}}(t)=\int\limits_{M_{low}}^{M_{max}}
  {\psi\left(t-\tau_{_M}\right) \varphi\left(M,t-\tau_{_M}\right)dM},
\end{equation}
where $M_{low}=8 M_\odot$ is the minimum mass of a presupernova. The
evolution of the chemical composition of a GC is described by the
equation of mass conservation for a particular element:
\begin{eqnarray}
&&  \frac{d}{dt}\left(Z_i M_g\right)=\nonumber\\
&&  -Z_i\left(t\right)\psi\left(t\right)+
  \int\limits_{M_{min}}^{2M_{Ia}^{low}}
  {\psi\left(t-\tau_{_M}\right)\varphi\left(M,t-\tau_{_M}\right)X_i\left(M,t-\tau_{_M}\right)dM}+\nonumber\\
&&  +\left(1-\beta\right) \int\limits_{2M_{Ia}^{low}}^{2M_{Ia}^{up}}
  {\psi\left(t-\tau_{_M}\right)\varphi\left(M,t-\tau_{_M}\right)X_i\left(M,t-\tau_{_M}\right)dM}+\label{eq3}\\
&&  +\beta\int\limits_{2M_{Ia}^{low}}^{2M_{Ia}^{up}}
  {\left[\int\limits_{\mu_{min}}^{1/2}
    {\psi\left(t-\tau_{_{\mu M}}\right) \varphi\left(M,t-\tau_{_{\mu M}}\right) X_i^{Ia}\left(M,t-\tau_{_{\mu M}}\right)
     f\left(\mu\right) d\mu}
  \right]dM}+\nonumber\\
&&  +\int\limits_{2M_{Ia}^{up}}^{M_{max}}
  {\psi\left(t-\tau_{_M}\right) \varphi\left(M,t-\tau_{_M}\right) X_i\left(M,t-\tau_{_M}\right)
  dM}+\nonumber\\
&&  +\dot{M}_{Z_i}^{in}-\dot{M}_{Z_i}^{out},\nonumber
\end{eqnarray}
where $Z_i(t)$ is the relative abundance of element $i$
at time $t$; $X_i(M,t-\tau_{_M})$ and
$X_i^{Ia}(M,t-\tau_{_M})$ are the masses of element $i$
ejected by evolved stars and by binaries of mass $M$ that were born
at time $t-\tau_{_M}$; $M_{Z_i}^{out}$ and $M_{Z_i}^{in}$ describe
the exchange of heavy elements with the intragalactic medium through wind
and accretion, respectively; $M_{Ia}^{low}=1.5 M_\odot$ and
$M_{Ia}^{up}=8 M_\odot$ are the minimum and maximum masses of stars
in binaries; $\beta$ is the fraction of binaries in the mass
interval $2M_{Ia}^{low}$--$2M_{Ia}^{up}$ ; and $f(\mu)$
is the mass distribution of stars in binaries (where $\mu$ is the
ratio of the mass of the lighter component and the mass of the
binary, which varies from
$\mu_{min}=\max(M_{Ia}^{low}/M,1-M_{Ia}^{up}/M)$ to $0.5$).

The first term in~(\ref{eq3}) describes the decrease of
$Z_i$ in the interstellar medium due to star formation; the second
term corresponds to enrichment of the interstellar medium with $i$th
element by stars with masses in the range $\left(M_{min},2M_{Ia}^{low}\right)$; the third
term describes the contribution from single stars with
masses $M=(2M_{Ia}^{low},2M_{Ia}^{up})$; the fourth term
describes the input from Type Ia supernovae in binaries
with masses ranging from $2M_{Ia}^{low}$ to $2M_{Ia}^{up}$; the fifth term
is the contribution to $Z_i$ due to stars
more massive than $2M_{Ia}^{up}$; the last two terms describe the
exchange of heavy elements between the cluster and the external medium.

In the present study, the value of $X_i\left(M,t-\tau_{_M}\right)$
for stars with masses $M<M_{low}$ is determined by equation
\begin{equation}
  X_i\left(M,t-\tau_{_M}\right)=Z_i\left(t-\tau_{_M}\right)\left(M-M_r\left(M\right)\right).
\end{equation}
The yields of heavy elements by Type II supernovae we computed with using
the results of Woosley and Weaver~\cite{WoosleyWeaver1995}, while
for Type Ia supernovae we used
the yields from~\cite{Nomoto1997}. The latter yields are valid for solar
metallicity only, but the yields of Type Ia supernovae do not depend
substantially on the metallicity; the mass distribution of stars in binaries we took in the form~\cite{GreggioRenzini1983}:
\begin{equation}
  f\left(\mu\right) = 24\mu^2
\end{equation}

\subsection {Photometry}
At time $t$ the system retains the following number of stars with
the masses from $M$ to $M + dM$ born at time from $t'$ to
$t'+dt'$:

\begin{equation}
  dN = \Psi(t') \varphi(M,t') \Theta(\tau_{_M}-\tau) dM dt',\label{chislo}
\end{equation}
where $\Theta(x)$ is the Heaviside function. For a given stellar
mass and age $\tau=t-t'$, the stellar luminosity $L$ and effective
temperature $T$ can be found from a database of evolutionary tracks.
For this purpose we used the Padova data
base~\cite{Padova1,Padova2,Padova3,Padova4,Padova5} supplemented
with data for low mass stars from~\cite{VanDerBergh}.
For the color-magnitude diagram we have utilized the
widely used library of stellar spectra described
in~\cite{AA97,AA98} with the data for stars of effective
temperatures $T < 50 000 K$, and the supplementary library~\cite{Clegg}
with the data for hot stars $T > 50 000 K$.
If one defines the spectrum of a star as
\begin{equation}
  F(\lambda;L,T,g)=L\cdot f(\lambda;T,g),\label{svet}
\end{equation}
where $L$ is the luminosity in solar units, $f(\lambda;T,g)$,~the
spectrum normalized to the solar spectrum, and $g$, the
surface gravity, the absolute magnitude is given by
\begin{eqnarray}
  m&=&-2.5\lg\int\limits_0^\infty F(\lambda) \Phi(\lambda)
  d\lambda + m_{\rm calib}\nonumber\\
  &=&-2.5\lg L-2.5\lg\int\limits_0^\infty f(\lambda)
  \Phi(\lambda) d\lambda + m_{\rm calib},\label{velichina}
\end{eqnarray}
where $m_{\rm calib}$ is the calibration constant of the filter
corresponding to the given waveband, with the filter sensitivity
curve $\Phi(\lambda)$~\cite{bruz}.

\section{RESULTS OF MODELING}
Earlier~\cite{AJ} we have considered and theoretically motivated
a model of chemical evolution of GCs within the hypothesis of  self-enrichment with a varying IMF: the initial stage of star
formation was
characterized by a ``top-heavy'' IMF shifted to $M>8 M_\odot$, while the
transition to a second stage of star formation with a ``normal'' IMF
occurred after heavy elements started to dominate in radiative
cooling of the proto-cluster gas. Note, that the possibility for the first stars forming of pristine matter to have predominantly high masses is
confirmed recently in numerical experiments (see,~\cite{abel}).

Here, we present the results of computations of color-magnitude
diagrams $V$-($B$-$V$) for a similar model. Figure 1 shows the
$V$-($B$-$V$) plane for a GC model with the mass of
$M=10^6M_\odot$ that undergoes two episodes of star formation: the first
with
a Salpeter-like IMF, but shifted to higher masses, so that the
minimum mass of the stars formed is $M_{min}=8 M_\odot$; the maximum
mass of stars is assumed $M_{max}=100
M_\odot$ for all cases. The SFE in the first stage is taken $f_1=2.6\cdot
10^5$~cm$^3$g$^{-1}$s$^{-1}$, and the transition time to the second
stage (with a normal IMF) is $t=10$ Myr. We assumed for the second
episode of star formation $M_{min}=0.1 M_\odot$ and $f_2=2\cdot
10^7$~cm$^3$g$^{-1}$s$^{-1}$. In order to avoid overproduction of heavy
elements in supernovae explosions, we assumed that 95\% of the matter
ejected by supernovae is swept up after $t=10$ Myr.
In these assumptions two main sequence turnoffs appear clearly in the
color-magnitude diagram.

\begin{figure}
  \centering
  \includegraphics[width=120mm]{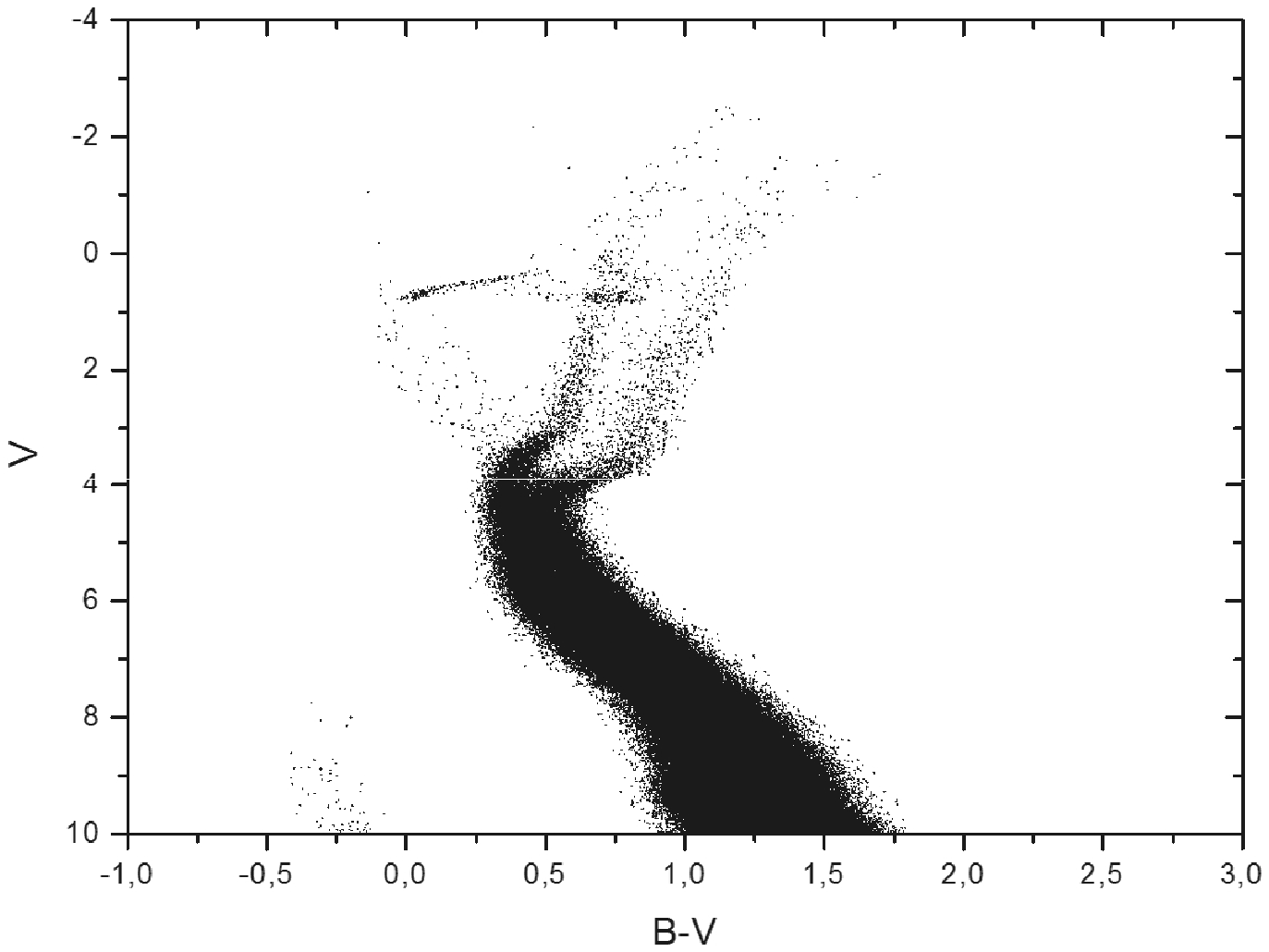}
  \caption{\label{ris1} Color-magnitude diagram for bimodal star formation. The initial
star formation episode has $M_{min}=8~M_\odot$ and $f_1=2.6\times
10^5$~sm$^3$g${-1}$s$^{-1}$, and lasts for $10^7$~yr; the second
star formation episode has $M_{min}=0.1~M_\odot$ and $f_2=2\times
10^7$~sm$^3$g$^{-1}$s$^{-1}$. At $t=10^7$~yr, 95\% of the matter
ejected by supernovae has left the cluster. }
\end{figure}

With the model parameters shown above, the presence of several turnoffs
is a natural consequence of the evolutionary processes of the
stellar system, if transformation of the IMF occurs as assumed in ~\cite{AJ} when the metallicity exceeds some critical value:
during the transition to the standard IMF the SFR
attains the peak value $\psi\sim 0.6$~$M_\odot$/yr due to
increase in the SFE by two orders of magnitude. The SFR remains near
its maximum level for 1~Myr. During this time, a significant
fraction of stars ($\sim 30\%$) form from gas that has already been
enriched by the first supernovae. The stars have formed in this
period have metallicities $[Z]\simeq -2$ and concentrate mainly
toward the left side of the diagram. A fraction of them fall by the
present into the region of the left subgiant branch, as seen in
Fig.~\ref{ris1}. After the transition to the ``standard'' IMF,
the metallicity of the stars in the system continues to increase. At
time $\sim 2\cdot 10^8$~yr, the metallicity attains a practically
constant asymptotic value $[Z]\sim -0.8$. In further continuing
star formation, an additionanl population of stars is born in the
cluster, now with a higher metallicity, close to the asymptotic
value ($\sim 10\%$).

\begin{figure}
  \centering
  \includegraphics[width=120mm]{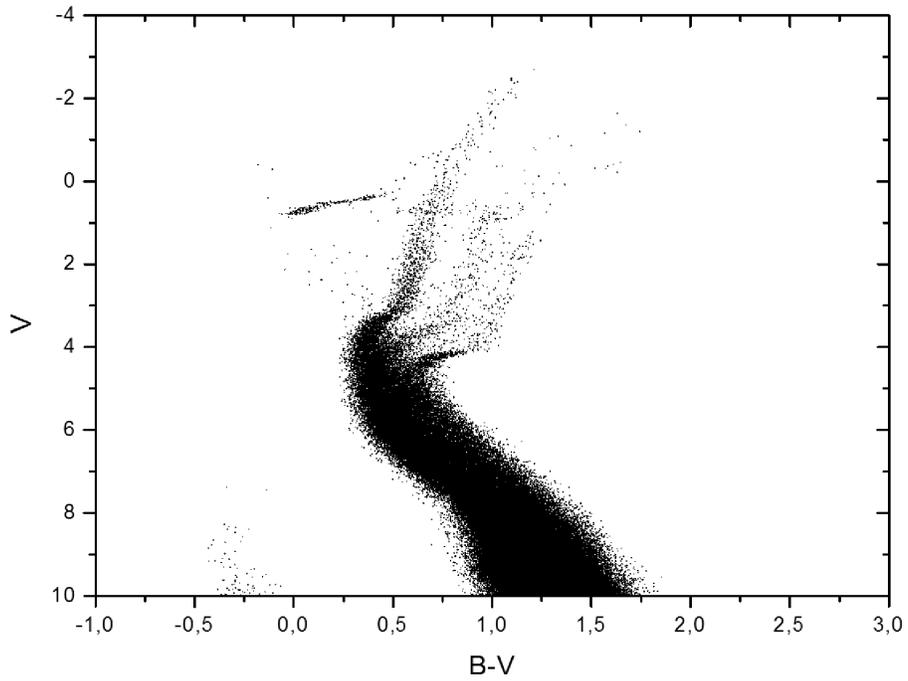}
  \caption{\label{ris2} Same as in
Fig. 1 for $f_2=2\times 10^8$ sm$^3$g$^{-1}$s$^{-1}$. }
\end{figure}

By varying the free parameters of the model, such as
the time for transition to the standard IMF,
star formation efficiencies in the first ($f_1$) and second ($f_2$)
episodes, the time when the matter from supernovae explosions begins
to leave the system, and the fraction of lost matter, one can
obtain even more complex color-magnitude diagrams
in this scenario. An example
is shown in Fig.~\ref{ris2}, which corresponds to
SFE on the second stage an order of magnitude higher than
before ($f_2=2\cdot 10^8$~sm$^3$g$^{-1}$s$^{-1}$), with all other
parameters fixed. We stress that the transition from the
diagram with two turnoff points to the one with three turnoff points
is reached in this case by increasing only one parameter ~---
$f_2$~--- without necessty to regulate ``by hands''
the sequence of several
star formation bursts. In turn, the SFE may be determined by a
number of factors, which at present cannot be estimated confidently.
For instance, in a model in which the star
formation is self-regulated by ionizing stellar
radiation~\cite{cox}, the coefficient $f_2$ may be critically determined
by inhomogeneity of the interstellar gas after the
first burst of star formation and the fraction of ionizing radiation
able to escape the proto-cluster.

An increase in the SFE results in a higher than in the latter case
number of stars in the left side of the diagram. On the other hand,
it leads to a faster depletion of gas, and therefore to a higher
enrichment in metals ($[Z]\sim -0.3$).
A fraction of the stellar population born
during the period when the metallicity of the system is maximum reaches
currently the extreme right of the subgiant branch.
Later on, the gas metallicity decreases due to mass loss of
evolved low- and intermediate-mass stars, and attains a constant
level, corresponding to the one of stars currently located in the
middle subgiant branch.

In the case considered here the presence of several turnoff points is a
consequence of natural evolution of the stellar system, and does
not require stronger assumptions about the nature of the star
formation, or the existence of a stellar population with
unusually high helium abundance ($Y\sim~0.4$~\cite{B04, Norris2004,
P05, Lee2005}), which would require an increase in the metal
abundance by $\Delta Z \sim \Delta Y/5 \sim 0.08$.
The diagrams in Figs.~\ref{ris1} and ~\ref{ris2} are plotted with
accounting the observational errors. We used an approximation of
the data from~\cite{site} to calculate the magnitude dependence of
the rms errors.

\begin{figure}
  \centering
  \includegraphics[width=120mm]{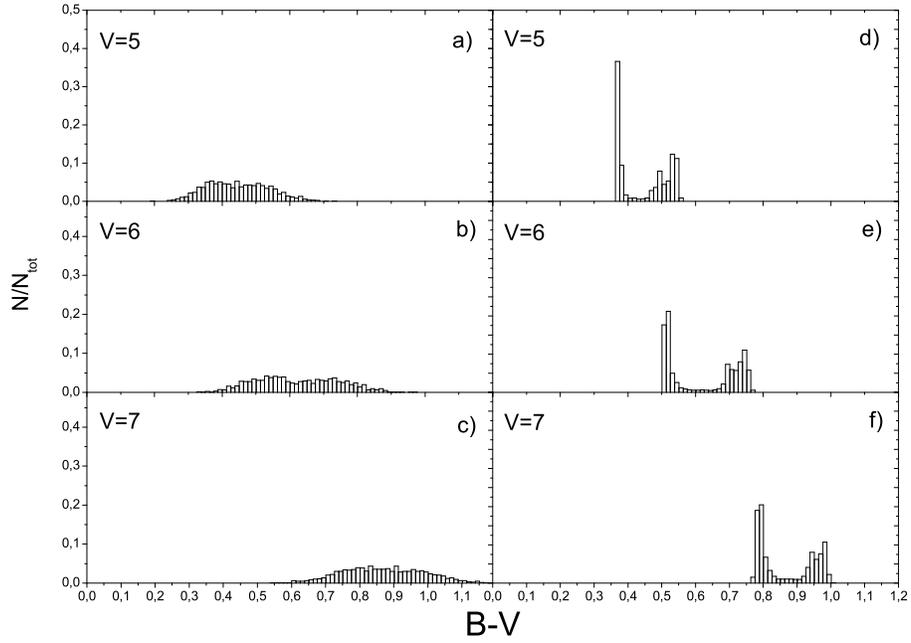}
  \caption{\label{ris3} Number of
stars as a function of the color corresponding to cuts of the
color-magnitude diagram shown in Fig. 1 for $V=5,~ 6,~ 7$. Diagrams
(a), (b), (c) correspond to the plotted histogram, taking into
account the observational errors, while diagrams (d), (e), (f)
correspond to the plotted histogram without accounting observational
errors.}
\end{figure}

\begin{figure}
  \centering
  \includegraphics[width=120mm]{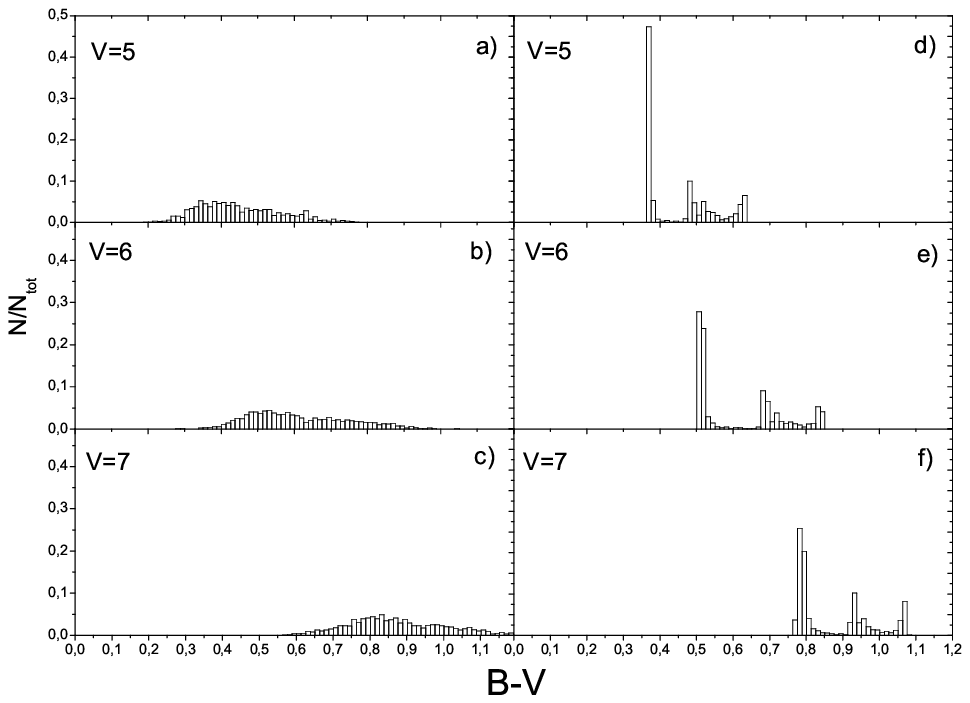}
  \caption{\label{ris4} Same as in
Fig. 3 for analogous cuts of the color-magnitude diagram presented
in Fig. 2.}
\end{figure}

The presence of several turnoff points indicates also that the main sequence itself has a complex structure. However, it is difficult to
distinguish this structure from such kind of color-magnitude diagrams,
mainly because of large observational errors for low-mass stars. In order
to analyze the main sequence, we calculated several color distributions
of stars along the ``cuts'' corresponding to a fixed value of V as suggested in~\cite{Helium}.
Figures~\ref{ris3} and ~\ref{ris4} show the dependences of the
relative numbers of stars for three values of $V$.
Figure~\ref{ris3} corresponds to the color-magnitude diagram shown
in Fig.~\ref{ris1} and Fig.~\ref{ris4} to the diagram shown in
Fig.~\ref{ris2}. In these figures left panels show the
distributions accounting the Gaussian smearing described
above; for comparison, we show in the right panels the
distributions obtained for the same model without contribution from
observational errors. These figures show that even though the system does
possess several main sequences, the distinction between them will not always be revealed in the observational color-magnitude diagrams.

A group of stars located on the continuing of the main sequence to
$V\sim 1$, in the region occupied by so-called blue stragglers, is
clearly seen. In our models, these are the youngest stars, born
in late stages of evolution of the GC. For instance, the group of
stars in Fig.~\ref{ris2} that can be identified with blue stragglers
is characterized by lower metallicities ($[Z]\sim -1$) than the
stars located in the right side of the diagram ($[Z]\sim -0.3$).
This relatively small group of blue stragglers is born in the model
shown in Fig.~\ref{ris2} at late evolutionary stages of stellar
system when the metallicity of the gas begins to decline due to
mass loss by low-mass stars with lower metallicities. Since
the gas mass remained in the system by this time is small, the
mass income with lower metal abundances becomes
sufficient to significantly decrease the gas metallicity~\cite{AJ}.
This identification is not acceptable in reality, however, since
blue stragglers typically have anomalous chemical composition,
which probably indicate their connection to active mass exchange in
binaries. Note that there are no stars in the region of the blue
stragglers in the color-magnitude diagrams if the model assumes
that star formation does not occur at ages exceeding 1~Gyr.


\subsection {A Model with Pre-Enrichment}
Models with self-enrichment assume a longer evolution of the system
than models with pre-enrichment. Indeed, models with pre-enrichment
suggest that the bulk of metals has been already injected into
the gas by some external sources on the proto-cluster stage.
Therefore, over the entire subsequent cluster evolution only a slight  enrichment is allowed, which does not
significantly exceed the amount of the pre-enrichment.
In turn, this means
that in models with self-enrichment star formation has to occur in a
mode that prevents the birth of stars from newly enriched matter. In
other words, in such models, newly enriched matter must be swept out
of the cluster before it will be able to cool radiatively and give
rise to new generations of stars. A sufficient condition for
such a mode to occur is a rapid star formation on early stages,
which results after several Myr in a large energy release enough to
prevent the enriched gas from cooling. A rough estimate of
this condition can be obtained by assuming that the
supernovae energy rate exceeds the total rate of radiation losses
in the gas:

\begin{equation}
\label{uslov} \nu_{sn}E_{sn}>\Lambda(T)n^2V,
\end{equation}
where $\nu_{sn}$ and $E_{sn}=10^{51}$~erg are the supernova rate in
a GC, and the supernova explosion energy,
$\Lambda(T)\sim 3\cdot 10^{-23}$~erg~sm$^3$s$^{-1}$ is the rate of
radiative losses at gas temperatures behind the shock
$T\sim (1-3)\cdot 10^5$~K, $n$ is the number density of
the gas and $V$ is the volume of a GC. For $n\sim 10^3$~см$^{-3}$
and $R\sim 20$~pc, the estimate (\ref{uslov}) gives $\nu_{sn}\sim 3$
supernovae per year, which is equivalent to a SFE $f\sim 2\cdot
10^8$~sm$^3$g$^{-1}$s$^{-1}$ an order of magnitude higher than
commonly assumed value. This is obviously an overestimate, since the
binding energy of globular clusters is low, and a substantially lower energy
release is sufficient to remove the gas from the cluster. Indeed,
it is easy
to show that, even for the SFE assumed above, $f\sim 2\cdot
10^7$~sm$^3$g$^{-1}$s$^{-1}$, the number of supernovae that would
explode in the cluster over time sufficient for the volume filling
factor of the hot gas to equal one is
$N_{sn}\sim 600$, whose total energy release
exceeds the binding energy of a cluster with mass $M=10^6 M_\odot$
and initial radius $R=20$~pc by two orders of magnitude. Below we will
adopt precisely this SFR for the model with pre-enrichment,
$f=2\cdot 10^7$~sm$^3$g$^{-1}$s$^{-1}$.

\begin{figure}
  \centering
  \includegraphics[width=120mm]{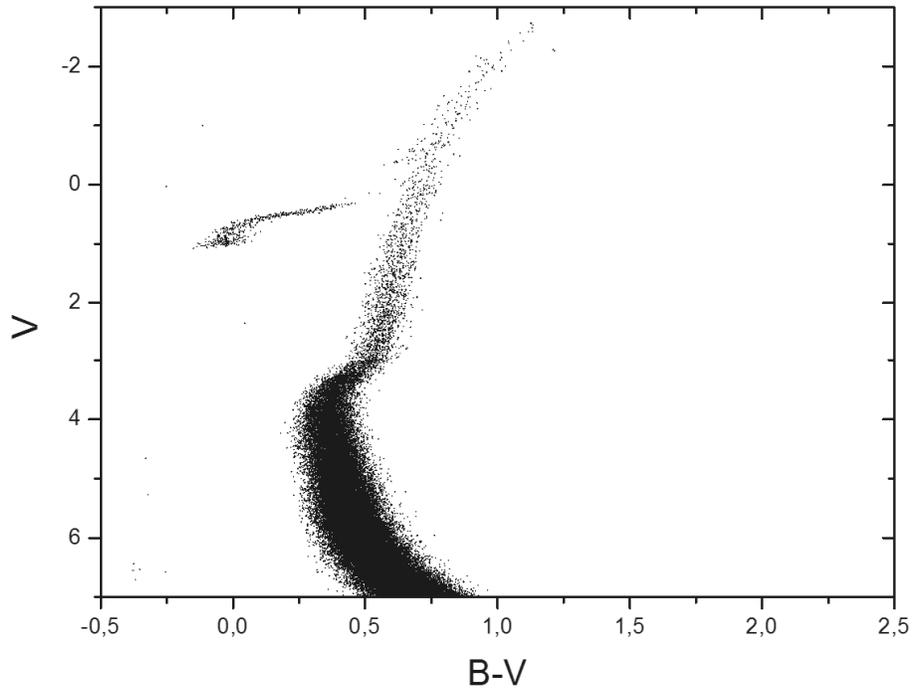}
  \caption{\label{ris5} Color-magnitude diagram for the model with primordial
self-enrichment $[Z]_{\rm in}=-1.5$; the SFR is specified to be zero
at $t=10^8$~yr. }
\end{figure}

Figure~\ref{ris5} shows the results of computing the color-magnitude
diagram for a GC with a mass of $M=10^6 M_\odot$ which is
pre-enriched to metallicity $[Z]=-1.5$ and begins its
evolution with radius $R=20$~pc and the SFE given above. Similar to
the previous figures, Fig.~\ref{ris5} is plotted with accounting a
Gaussian smearing. An important feature of this model is that, when
the first supernovae begin to eject heavy elements at about $t >
10^7$~ yr, the bulk of the matter with the initial chemical
abundances is already contained in stars. Thus if we assume that under
these circumstances the gas ejected by supernovae is easily removed
from the system, the number of stars with metallicities exceeding
the initial value will be negligibly small. Figure~\ref{ris5} shows
the results for a model GC in which practically all stars have the
initial metallicity.

\begin{figure}
  \centering
  \includegraphics[width=120mm]{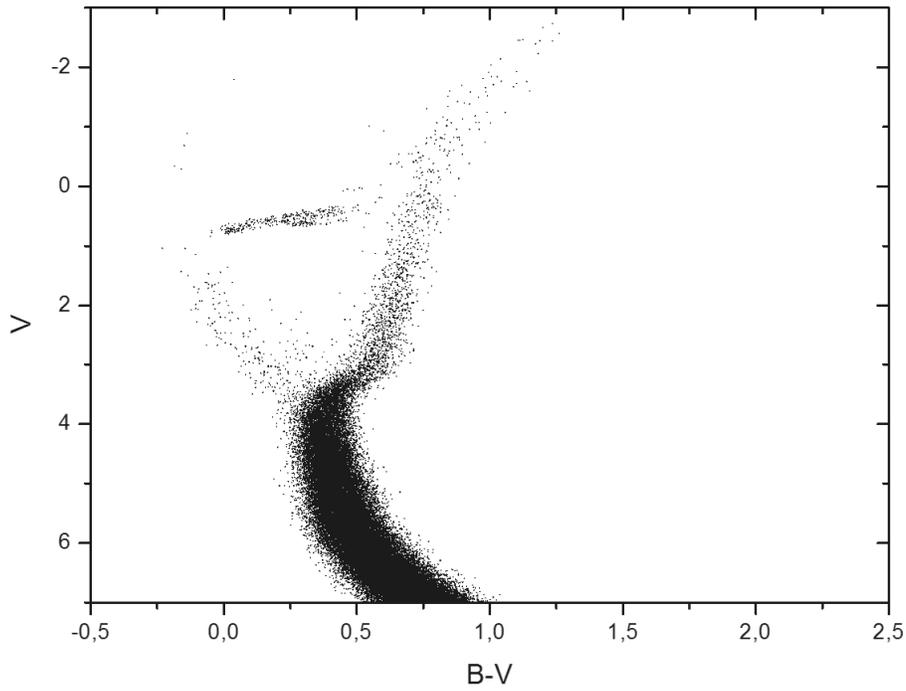}
  \caption{\label{ris6} Same as in
Fig. 1, but with the fraction of matter ejected by supernovae and
lost by the system at time $t=10^7$~yr being 99.7\%. }
\end{figure}

\begin{figure}
  \centering
  \includegraphics[width=120mm]{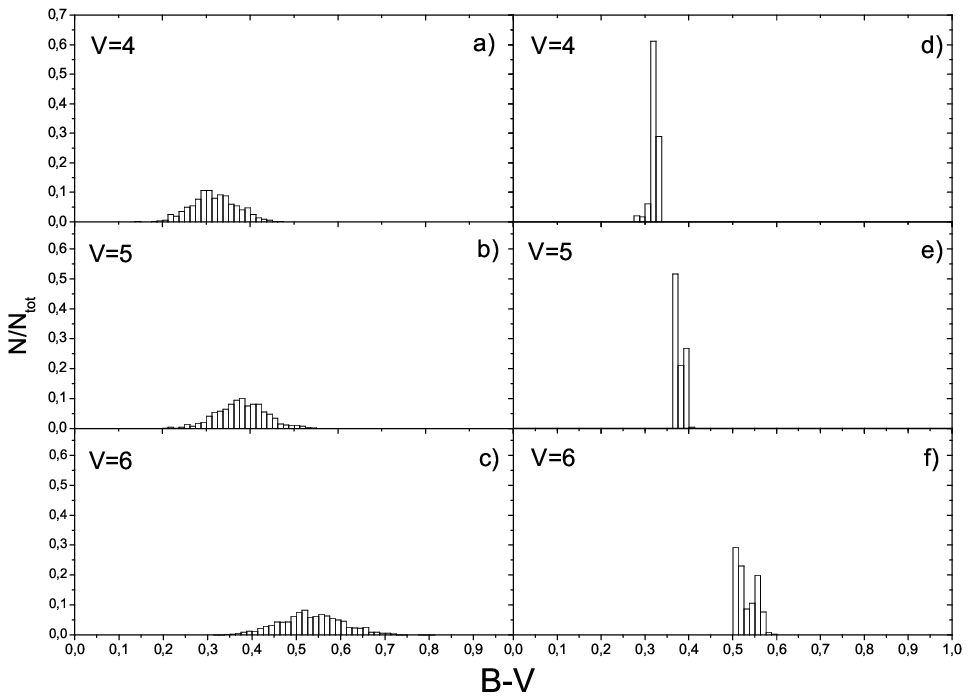}
  \caption{\label{ris7} Same as in
Fig. 3 for the analogous cuts of the color-magnitude diagram
presented in~Fig.~7.}
\end{figure}

It is obvious that a similar diagram can also be obtained in
a model with self-enrichment if the parameters of star
formation are such that the separation (in $B-V$) between the main
sequences is smaller than the observational errors. In this case, it
is practically impossible to distinguish between
the two populations on observational color-magnitude diagrams. As an illustration of this
situation, Figures~\ref{ris6} and~\ref{ris7} present a
color-magnitude diagram and color distributions of relative numbers
of stars, respectively. In this model, the first burst of star
formation has $M_{min}=8 M_\odot$, $M_{max}=100 M_\odot$, and
$f_1=2.6\cdot 10^5$~sm$^3$g$^{-1}$s$^{-1}$; the transition
to the second burst with a ``standard'' IMF occurs at $t=10$~Myr; the
second burst of star formation has $M_{min}=8 M_\odot$, $M_{max}=100
M_\odot$, and $f_2=2\cdot 10^7$~sm$^3$g$^{-1}$s$^{-1}$, with 99.7\%
of the matter ejected by supernovae swept out of the cluster after
$t=10$~Myr. In this case, the absence of multiple turnoff points is
a consequence of a larger fraction mass loss
than in the model shown in Fig.~\ref{ris1}. Note that the
model computation produces two stellar populations which correspond
to low metallicities $[Z]=-2$ and -1.6; this makes it even more
difficult to distinguish between the populations, due to the proximity of
the isochrones for low metallicities.

Thus, the absence of multiple turnoff points in observational
color-magnitude diagram does not necessarily mean that matter of
the corresponding GC was pre-enriched in heavy elements,
since the separation that arises in models with self-enrichment may
be smaller than the errors in the colors. In
self-enrichment scenarios, it may be easier to distinguish stars
with different metallicities within more massive GCs. In fact, we
would expect a decrease of the metallicity of the main fraction of
stars with increasing cluster mass~\cite{MZcor,AJ}. At the same
time the later generation of stars has as a rule metallicities
close to the asymptotic value $[Z]\sim -0.8$, which only weakly depends
on the cluster mass. Observing this separation for GCs
with masses of $M\sim 10^5~M_\odot$ requires an increased
accuracy for magnitudes of $\Delta m\sim 0.03-
0.07$. The proposed interpretation of the color-magnitude diagram
of $\omega$ Cen as having multiple main sequence turnoff points
with several distinct stellar populations is justified by
spectroscopic measurements of metallicities in $\omega$ Cen.
Several main
sequences are also observed in NGC 2808 [43]. With accounting the fact that spectroscopic metallicities of
different branches of the main sequence in NGC 2808 are nearly equal, Piotto et al.~\cite{P07} suggest a scenario where the multiplicity of the main sequence is connected with different helium abundances ($Y \sim 0.24-0.4$). Thus, the two cases~--- $\omega$ Cen and NGC 2808~--- seem
to represent two possible scenarios for the evolution of GCs: one
produces several populations with different metallicities, while
another produces populations with similar metallicities but very
different helium abundances. From this point of view in order to
clarify the origin of this multiplicity, detection
of multiple main sequence turnoff points in other Galactic GCs and
a precise spectroscopic determinations of the iron abundance will
play crucial role.

\section{CONCLUSIONS}
\noindent

We have applied a simple single-zone model to analyze color-magnitude
diagrams for globular clusters. Our results show the following.

1. The presence of two or more main sequence turnoff points in the
color-magnitude diagrams of GCs can be naturally explained in models
with self-enrichment, with an initial stellar mass function that is
initially shifted toward high masses, but then changes to a normal
IMF.

2. This represents an argument in favor of the self-enrichment of
GCs with multiple turnoffs (such as $\omega$ Cen) in metals due to
internal processes within the GC.

3. On the other hand, the uncertainties in the luminosities and
magnitudes of stars in GCs are still fairly large, and do not always
enable resolution of the main sequences in color-magnitude diagrams
into subgroups. In other words, the absence of multiple turnoff
points does not necessarily require an external (pre-)
enrichment in metals.

\section{ASKNOWLEDGMENTS}
\noindent The authors thank V. I. Korchagin for discussions.
This study was supported by the Russian Foundation for
Basic Research (project code 06-02-16819) and the State Education
Agency (project code RNP 2.1.1.3483).





\newpage



\end{document}